\documentstyle[aps,prd,preprint]{revtex}
\begin{document}
\tightenlines

\def\stacksymbols #1#2#3#4{\def\theguybelow{#2}
    \def\verticalposition{\lower#3pt}
    \def\spacingwithinsymbol{\baselineskip0pt\lineskip#4pt}
    \mathrel{\mathpalette\intermediary#1}}
\def\intermediary#1#2{\verticalposition\vbox{\spacingwithinsymbol
      \everycr={}\tabskip0pt
      \halign{$\mathsurround0pt#1\hfil##\hfil$\crcr#2\crcr
               \theguybelow\crcr}}}
\def\lapproxeq{\stacksymbols{<}{\sim}{2.5}{.2}}
\def\gapproxeq{\stacksymbols{>}{\sim}{3}{.5}}

\title{Singular Instantons and Creation of Open Universes}
\author{Alexander Vilenkin\footnote{Electronic address: vilenkin@cosmos2.phy.tufts.edu}}

\address{Institute of Cosmology,
        Department of Physics and Astronomy,\\
        Tufts University,
        Medford, Massachusetts 02155, USA}
\date{\today}
\maketitle

\begin{abstract}
Singular instantons of the type introduced by Hawking and Turok lead to
unacceptable physical consequences and cannot, therefore, be used to
describe the creation of open universes.
\end{abstract}

\newpage
In a recent paper \cite{HT}, Hawking and Turok presented an instanton
solution which they interpret as describing the creation of an open 
inflationary universe.
An unusual feature of this instanton is that it is singular : both the
curvature and the scalar field diverge at one point.
Hawking and Turok argue that it is nonetheless a legitimate instanton,
since the singularity is integrable and the Euclidean action is finite.
I do not find this argument very convincing.
In a singular instanton, the field equations are not satisfied at the
singularity. Such an instanton is not, therefore, a stationary point
of the Euclidean action, and it is not at all clear that it will give 
a dominant contribution to the Euclidean path integral.
Moreover, if the Hawking-Turok(HT) instanton is allowed, we will then have 
to admit a host of other instantons with integrable singularities.
Here, I shall give an example of such an instanton and argue that
it leads to physically unacceptable consequences.

Let us consider the model of a massless scalar field $\phi$
interacting with gravity. The corresponding Euclidean action 
is\footnote{A massless field has been chosen for simplicity.
A similar instanton can easily be constructed with a massive scalar 
field.}

\begin{equation}
	S_E = \int d^4x \sqrt{g} \left[ -{R \over 16\pi G}
		+ {1 \over 2}(\partial\phi)^2 \right ] 
		+ S_{boundary}\,.
\label{eq:action}
\end{equation}
As in \cite{HT}, we shall be interested in an O(4)-symmetric instanton
described by the metric

\begin{equation}
	ds^2 = d\sigma^2 + b^2(\sigma) (d\psi^2 
		+ \sin^2\psi d\Omega_2^2)
\label{eq:metric}
\end{equation}
with $\phi = \phi(\sigma)$.
The field equations for $b(\sigma)$ and $\phi(\sigma)$ are

\begin{equation}
	\phi'' + 3{b' \over b}\phi' = 0\,,
\label{eq:phieq}
\end{equation}
\begin{equation}
	b'' = -{8\pi G \over 3}b\phi'^2\,,
\label{eq:beq}
\end{equation}
where primes stand for derivatives with respect to $\sigma$.
We choose the boundary conditions corresponding to an 
asymptotically-flat instanton,

\begin{equation}
b(\sigma) \approx \sigma\,,\qquad
\phi(\sigma) \to 0 \qquad (\sigma\to \infty)\,.
\label{eq:bc}
\end{equation}
>From (\ref{eq:phieq}) it follows that

\begin{equation}
\phi' = -{C \over b^3}\,,
\label{eq:phiprime}
\end{equation}
where $C=const.$
Substituting this in (\ref{eq:beq}) and integrating, we have

\begin{equation}
b'^2 - {4\pi GC^2 \over 3b^4} = 1\,,
\label{eq:bprime}
\end{equation}
where the constant of integration has been chosen according to
the boundary condition (\ref{eq:bc}).
The asymptotic forms of the solution at large and small $\sigma$
are easily found.
At $\sigma\to\infty$,

\begin{equation}
b(\sigma) \approx \sigma + {\cal O}(\sigma^{-3})\,,
\label{eq:binfty}
\end{equation}
\begin{equation}
\phi(\sigma) \approx C/2\sigma^2\,.
\label{eq:phiinfty}
\end{equation}
As we move towards smaller values of $\sigma$, $b(\sigma)$ decreases
and $\phi(\sigma)$ grows until $b$ vanishes and $\phi$ diverges
at some $\sigma = \sigma_* \sim (C/m_p)^{1/2}$.
The form of the solution near $\sigma_*$ is given by

\begin{equation}
b(\sigma) \approx (12\pi GC^2)^{1/6}(\sigma-\sigma_*)^{1/3}\,,
\label{eq:bstar}
\end{equation}
\begin{equation}
\phi(\sigma) \approx (12\pi G)^{-1/2}\ln(\sigma-\sigma_*) + const\,.
\label{eq:phistar}
\end{equation}
The transition between the two regimes occurs at 
$\sigma\sim (C/m_p)^{1/2}$, $\phi\sim m_p$, where $m_p$ is the Planck
mass.
The singular behavior (\ref{eq:bstar}), (\ref{eq:phistar}) is 
identical to that of HT instanton.
So if HT instanton is legitimate, then we have no reason to reject
the asymptotically-flat instanton described above.

To evaluate the Euclidean action of our instanton, we first note that 
the trace of Euclidean Einstein equations reads 
$R = 8\pi G(\partial\phi)^2$. 
Hence the two terms in the square brackets of Eq. (\ref{eq:action})
cancel, and the only contribution to the action comes from the
boundary term at $\sigma = \sigma_*$.
This term was omitted in Ref. \cite{HT} ; its presence was pointed
out to me by J. Garriga \cite{GG}. In general, the boundary term is 
given by \cite{GH}

\begin{equation}
S_{boundary} = -{1 \over 8\pi G}\partial_{normal}
({\rm Volume\; of\; boundary})\,.
\label{eq:ghbc}
\end{equation}
In our case, the volume of a 3-sphere $\sigma = const$ is
$2\pi^2 b^3(\sigma)$, and with the aid of (\ref{eq:bstar})
we obtain

\begin{equation}
S_E = S_{boundary} = \left( {3\pi^3 \over 4G} \right)^{1/2} C\,.
\label{eq:baction}
\end{equation}

Analytic continuation of the metric (\ref{eq:metric}) to Lorentzian
time, $\psi = \pi /2 + it$, gives \cite{HT}

\begin{equation}
ds^2 = d\sigma^2 + b^2(\sigma)(-dt^2 + \cosh^2t d\Omega_2^2)\,.
\label{eq:lormetric}
\end{equation}
It describes an asymptotically-flat spacetime with a singular 
hypersurface at $\sigma = \sigma_*$.
The singularity has the form of a sphere expanding at a speed close
to the speed of light. This can be seen by considering a sphere,
$b(\sigma)\cosh t = R$, of a constant proper radius
$R \gg (C/m_p)^{1/2}$.
It is easily verified that this sphere is engulfed by the singularity
in a finite proper time $\Delta\tau \approx R$ from the moment
of nucleation $t =0$.

If the singular Euclidean solution described above is accepted
as a legitimate instanton, we will have to conclude that
flat space is unstable with respect to nucleation of singular bubbles.
The nucleation probability is ${\cal P} \propto e^{-S_E}$
and is not exponentially suppressed for sufficiently small values
of $C \lapproxeq m_p^{-1}$.
If this picture were correct, most of the universe would have already
been overrun by expanding singular bubbles.
Since this picture is in glaring contradiction with observations,
we have to conclude that singular instantons of the Hawking-Turok
type cannot be used to describe quantum nucleation processes.
In particular, they cannot describe nucleation of open universes.

I am grateful to Jaume Garriga for very useful discussions during 
the course of this work.

\bigskip

{\it Note added:}

\medskip

Hawking and Turok responded to the e-print version of this paper by
pointing out that in their case the singularity never hits an observer
in the open universe spacetime region, while in the case of my
asymptotically flat instanton the singular bubble expands to engulf
the whole space \cite{HT2}.  It appears, however, that this
observation is beside the point.  The relevant question is: which
instantons contribute to the Euclidean path integral?  If singular
instantons do contribute, then my instanton should be included as
well, with all its unpleasant consequences.

\end{document}